\documentclass[pdflatex, sn-mathphys]{sn-jnl}

\usepackage[output-decimal-marker={.}]{siunitx}
\DeclareSIUnit{\belmilliwatt}{Bm}
\DeclareSIUnit{\dbm}{\deci\belmilliwatt}
\DeclareSIUnit{\samplepersecond}{Samples/s}
\DeclareSIUnit{\decade}{dec}
\DeclareSIUnit{\baud}{baud}

\jyear{2026}%

\begin{document}

\title[ ]{Sequential Monitoring and Control of a Silicon Photonic Coherent Beam Adder and Analyzer}

\author[1]{\fnm{Samuele} \sur{De Gaetano}} \email{samuele.degaetano@polimi.it}
\author[1]{\fnm{Monica} \sur{Crico}}
\author[2]{\fnm{Giorgio} \sur{Ferrari}}
\author[1]{\fnm{Marco} \sur{Sampietro}}
\author[1]{\fnm{Francesco} \sur{Morichetti}}
\author[1]{\fnm{Andrea} \sur{Melloni}}
\author[1]{\fnm{Francesco} \sur{Zanetto}}

\affil[1]{\orgdiv{Department of Electronics, Information and Bioengineering}, \orgname{Politecnico di Milano}, \orgaddress{\postcode{20133}, \state{Milano}, \country{Italy}}}

\affil[2]{\orgdiv{Department of Physics}, \orgname{Politecnico di Milano}, \orgaddress{\postcode{20133}, \state{Milano}, \country{Italy}}}

\abstract{Joint communication and sensing applications require devices that can analyze multiple electromagnetic waves and process them in real time directly in the analog domain. In optics, the growing maturity of photonic integrated platforms allows the fabrication of complex circuits that can perform such operations, but their large number of sensors and actuators requires scalable control strategies to efficiently monitor and actively stabilize their functionality at runtime. In this work, we report on a multi-aperture silicon photonic programmable circuit that operates both as a coherent beam adder and a multi-aperture beam analyzer. The circuit consists of a reconfigurable mesh of Mach-Zehnder interferometers controlled through monolithically integrated electronic circuits, which are used to serialize/deserialize the readout of integrated sensors and the driving of actuators with a time-multiplexed addressing scheme. The circuit operation is validated in a communication and sensing scenario, where the photonic chip is used to simultaneously receive a \SI{25}{\giga\bit\per\second} high-speed transmission and to measure the phase difference between the input light beams.}


\maketitle

\section{Introduction}\label{sec1}

\noindent Programmable photonic integrated circuits (PICs), like meshes of Mach-Zehnder interferometers (MZI) for optical computing or interconnect applications \cite{bogaerts_2020_programmable}, require an electronic control unit to configure, calibrate and manage them, monitor their functionality in real-time and keep it stable against environmental effects \cite{ padmaraju_2014_resolving,milanizadeh_2020_control,morichetti_2021_polarization,li_3d_2021,tan_2021_towards,jayatilleka_2019_photoconductive}. This feedback-based approach, which is often replaced by off-line calibration procedures and look-up tables, becomes essential for reliably operating large-scale circuits in realistic and in-the-field scenarios. In addition, in some applications, the use of pre-calculated look-up tables is simply not possible, such as in the case of adaptive PICs whose working point needs to be dynamically adjusted to match the time-varying characteristics of an input signal \cite{martinez_2024_selfadaptive,seyedinnavadeh_2024_determining}. Thus, to ensure full reconfiguration and real-time control, these PICs require multiple on-chip light sensors and actuators, which are used to close the external electronic configuration and stabilization feedback loops. The number of such elements is strictly related to the circuit size and can easily scale up to hundreds in complex PICs \cite{fowler_2022_integrated,wang_2022_adaptive_tdm}. Therefore, solutions to integrate electronic functionalities into the same photonic chip have been investigated to reduce the number of electrical connections between PIC and electronics as well as to reduce the assembly complexity, enabling the extension of the control paradigm to large-scale circuits \cite{zanetto_2023_unconventional, rakowski_2020_45nm}.

Among programmable and adaptive PICs, coherent adders based on self-configuring structures have gained popularity in recent years. They are usually implemented with meshes of MZIs arranged in diagonal, binary tree or other topologies \cite{miller_2013_self}. Their credentials have already been validated in several applications, including optical mode manipulation and unscrambling \cite{annoni_2017_unscrambling,seyedinnavadeh_2024_determining}, turbulence and wavefront distortion mitigation in free-space optical (FSO) transmission \cite{milanizadeh_2022_separating}, photonic computing \cite{pai_2023_experimentally} and blockchain \cite{pai_2023_experimental}. The versatility of these PICs also enables their use in applications where a combination of functionalities is required, such as for the so-called joint communication and sensing (JCAS) paradigm in which channel estimation, signal processing and acquisition of the environmental conditions are performed at the same time as optical transmission \cite{chow_2024_recent,he_2023_integrated,chen_2023_field}.  

Here, we present an MZI-based PIC that combines an optical switch matrix and an optical coherent adder, together with monolithically integrated control electronics. The chip, realized in a commercial silicon photonic foundry, is operated with a time-multiplexed feedback strategy, enabled by on-chip CMOS circuits. Two integrated electronic multiplexers, employed respectively for the readout of light sensors and for phase actuation tasks, significantly reduce the number of external connections while still maintaining feedback control over the photonic chip, all within a small footprint area. After discussing the chip architecture in Section \ref{sec3}, Section \ref{sec4} describes the time-multiplexed feedback strategy developed to efficiently control the functionality of the photonic chip. Section \ref{sec5} experimentally assesses the performance of the proposed approach in a \SI{25}{\giga\bit\per\second} transmission experiment, and demonstrates the use of the coherent adder in a sensing scenario, where the circuit is used to track the phase difference between the input optical beams. Section \ref{sec7} summarizes the main results of the work.

\section{Architecture of the electronic-photonic chip}\label{sec3}

\begin{figure*}[!b]
	\centering
    \includegraphics[width=\textwidth]{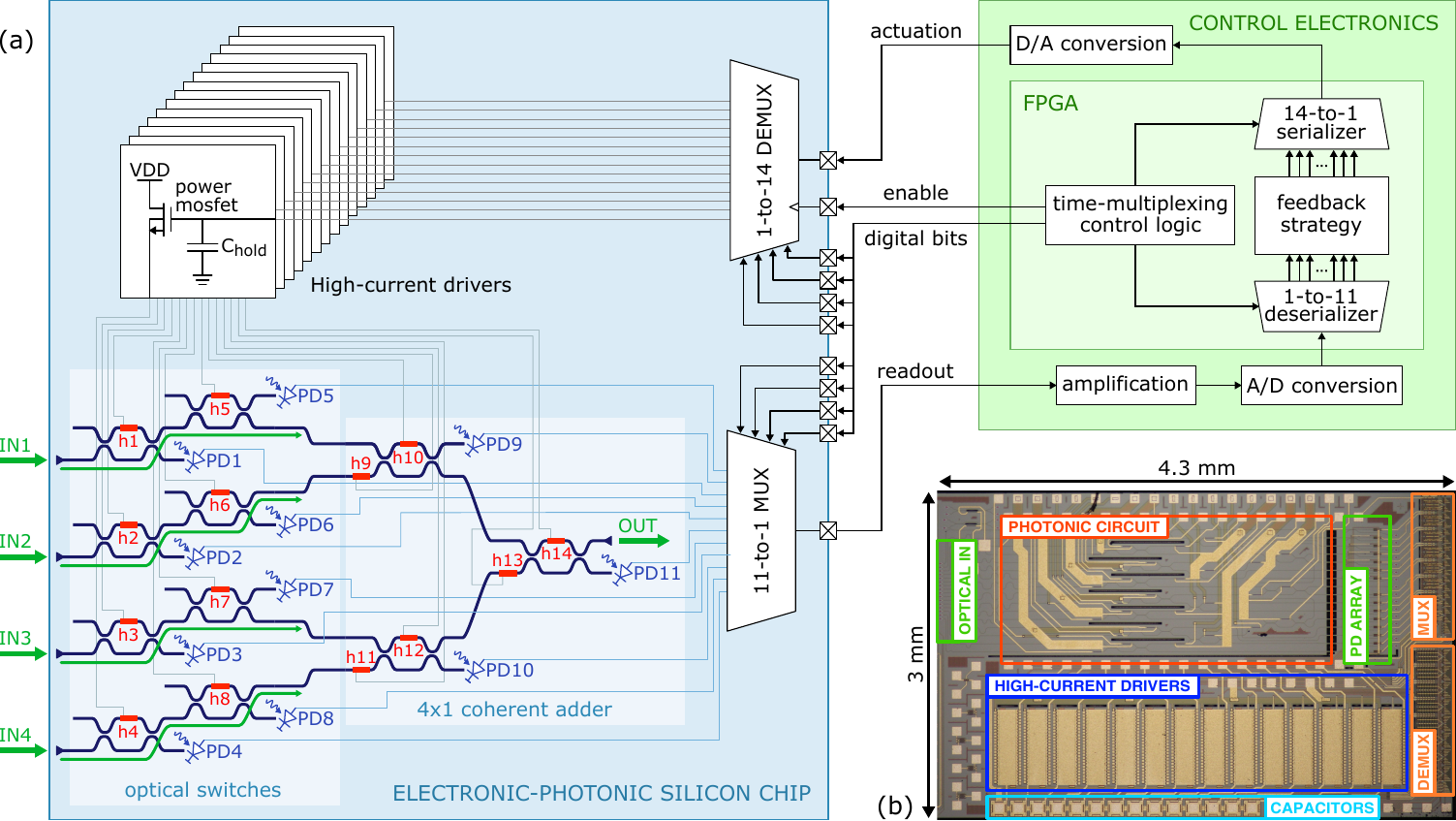}
	\caption{a) Schematic of the complete silicon chip (blue area), connected to the external control electronics (green area). The electrical signals needed to read the PDs and drive the heaters are serialized and deserialized with on-chip electronic circuits. b) Microscope photograph of the chip, highlighting the electronic and photonic sections. The chip has been manufactured by a commercial Silicon Photonics foundry \cite{AMF}, without modifications to the conventional fabrication process.}
	\label{fig1}
\end{figure*}

The architecture of the electronic-photonic chip is illustrated in Figure \ref{fig1}a. The entire circuit has a footprint of $\SI{4.3}{\milli\meter} \times \SI{3}{\milli\meter}$ (blue area) and is connected to an external digital electronic control board (green area). A microscope photograph of the die, which was fabricated in a commerical silicon photonic foundry (Advanced Micro Foundry, Singapore \cite{AMF}), is shown in Figure \ref{fig1}b. The circuit features 11 thermally-tunable balanced MZIs, which are arranged in a 2-stage switch matrix followed by a $4\times 1$ binary mesh topology that acts as a 4-input coherent adder \cite{miller_2013_self}. This architecture was chosen to demonstrate the proposed time-multiplexed control approach on both devices that are insensitive to the input phase (switches) and on phase-dependent components (coherent adder), thus providing a generalized validation. Each MZI is equipped with thermal phase shifters ($h1-h14$) and a dedicated on-chip germanium photodiode (PD) on the output drop port ($PD1-PD11$), allowing real-time monitoring and configuration of the photonic functionality. Surface grating couplers, designed for transverse-electric (TE) polarized light, are used to couple light in and out of the chip ports.  

To minimize the number of electrical input/output (I/O) connections required for interfacing the optical circuit with the external control board, an electronic multiplexer (11-to-1 MUX) has been integrated on the chip to sequentially route the current signals from the monitor PDs towards a single readout transimpedance amplifier and acquisition circuit. After digitization with an analog-to-digital converter (ADC), the serialized signals are separated into parallel processing chains inside the field-programmable gate array (FPGA) hosted in the external board that manages the control system. The information is digitally elaborated, serialized again and brought back into the analog domain with a single digital-to-analog converter (DAC), connected to a second integrated demultiplexer (1-to-14 DEMUX) that sequentially drives each MZI actuator. In this way, only two I/O signals are required to operate the whole optical circuit, in addition to the MUX/DEMUX address bits. The latter scale as $\log_2\left(N_{D}\right)$, where $N_{D}$ is the number of multiplexed devices (14 for the heaters, 11 for the integrated PDs). This approach becomes increasingly advantageous as the complexity of the PIC, and hence $N_D$, increases. 

While multiplexing the PD signals is relatively simple, the strategy to maintain the heaters working point stable when the DEMUX switches from one actuator to the next is more critical. To this end, we have integrated an electronic memory on the chip. Such memory is implemented with a sample-and-hold (S\&H) circuit connected to the high-current drivers needed to supply the thermal actuators \cite{zanetto_2023_time}, as shown in Fig. \ref{fig1}a. The S\&H circuit is constituted by the switches of the DEMUX together with a bank of on-chip memory capacitors ($\approx \SI{10}{\pico\farad}$), that store the heater voltage information and keep it stable when the DEMUX switches from one device to the next. This solution also relaxes the requirements of the external electronics. Indeed, a low-power external DAC is sufficient to charge and discharge the hold capacitors, while the electrical power required to operate the heaters is provided by the on-chip drivers. The latter are connected in a source-follower configuration to ensure a linear relationship between gate voltage and drain current \cite{zanetto_2023_time}.

Being digital circuits, the on-chip MUX/DEMUX are characterized by a negligible power consumption. Their static dissipation is $\approx\SI{200}{\nano\watt}$, while the dynamic contribution at the maximum switching frequency of $\SI{2}{\mega\hertz}$ remains below $\SI{100}{\micro\watt}$, about two orders of magnitude lower than the electrical power needed to operate a single heater \cite{zanetto_2023_unconventional}. The overall power budget of the on-chip electronics is therefore dominated by the actuators drivers, whose dissipation is intrinsically comparable to the electrical power delivered to the corresponding heaters.

Notice from Fig. \ref{fig1}b that half the chip area is occupied by the photonic circuit, including the I/O grating couplers, 30\% by the high-current drivers and hold capacitors and only the 9\% is dedicated to the electronic MUX and DEMUX. Thermal isolation trenches physically separate the electronic and photonic regions of the chip. These structures act as thermal barriers that mitigate thermal crosstalk between high-current drivers and optical waveguides, limiting unintended phase shifts induced by electronic power dissipation. The I/O electrical pads are visible in the lower left part of the chip and below the DEMUX. The other pads are used only for backup and testing purposes and are not needed to operate the chip.

\section{Time-multiplexed control algorithm}\label{sec4}

\begin{figure*}[!t]
	\centering
    \includegraphics[width=0.6\textwidth]{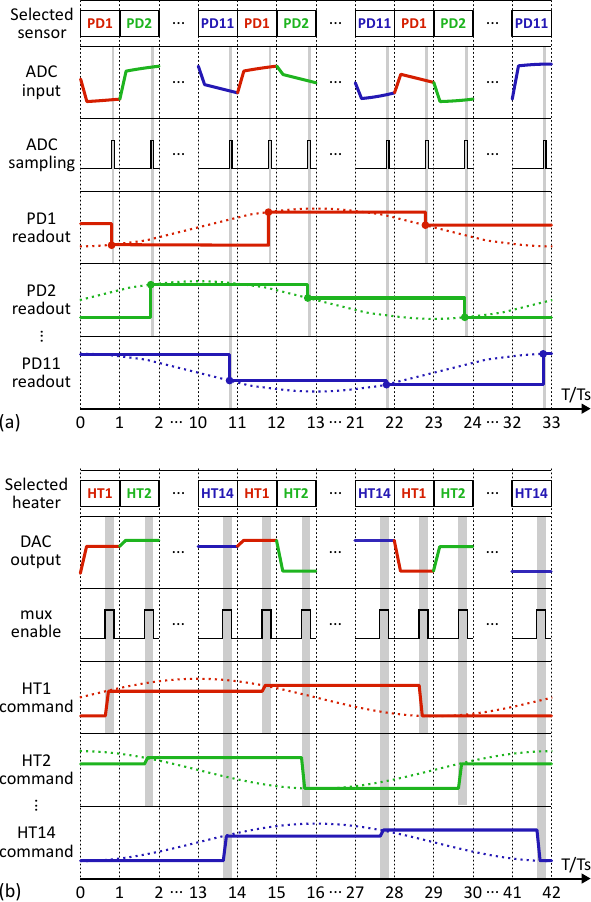}
	\caption{Time-multiplexed algorithm for sequential a) readout of multiple monitor PDs and b) driving of the on-chip thermal actuators. Reading and driving are cycled continuously and automatically.}
	\label{fig2}
\end{figure*}

Figure \ref{fig2}a shows in more detail the procedure for the sequential readout of monitor PDs with the integrated MUX. The time-multiplexing logic inside the FPGA automatically cycles the digital code that selects the PD to read. Each time this happens, the analog signal at the ADC input changes according to the sensor current. The initial transient observed in each time slot is due to anti-alias filtering and needs to be discarded to ensure a correct acquisition. To this end, the ADC sampling phase of the PD current (dashed curves in PD readout traces) is properly synchronized with the MUX switching. After digitization, the resulting bit stream is demultiplexed into 11 independent signals, each corresponding to the readout of a single PD, allowing parallel processing of the acquired information. The ADC sampling frequency is set to \SI{154}{\kilo\samplepersecond}, thus updating each channel at \SI{14}{\kilo\samplepersecond}. This rate is enough to monitor and counteract the typical perturbations that affect the operating point of each MZI. 

A similar approach is employed for the multiplexed driving of the actuators, as shown in Figure \ref{fig2}b. The command to each driver, coming from independent elaboration chains, is digitally serialized in the FPGA and generated by the single DAC on the electronic control board, with an update frequency of \SI{196}{\kilo\samplepersecond}. Due to unavoidable bandwidth limitations, the DAC output has a transitory phase when the voltage to be generated changes. Therefore, the sampling phase of each S\&H is suitably delayed with respect to the DAC update, thanks to a dedicated enable bit (mux enable) that turns on/off the DEMUX. This avoids unwanted instabilities in the actuators command voltage and results in a stable, non-volatile performance of the heater. The actuator to be operated in each time frame is addressed by properly selecting the DEMUX digital code and its command is kept constant till the next cycle period by the hold capacitor. Previous characterization of the S\&H device highlighted that no significant discharge of the hold capacitor is observed for refresh rates above \SI{10}{\hertz} \cite{zanetto_2023_time}. Therefore, the selected \SI{14}{\kilo\hertz} update frequency per actuator is enough to avoid any performance degradation.

\section{JCAS assessment of the coherent adder}\label{sec5}

\subsection{Experimental setup}\label{sec5_1}
\begin{figure*}[!t]
	\centering
\includegraphics[width=\linewidth]{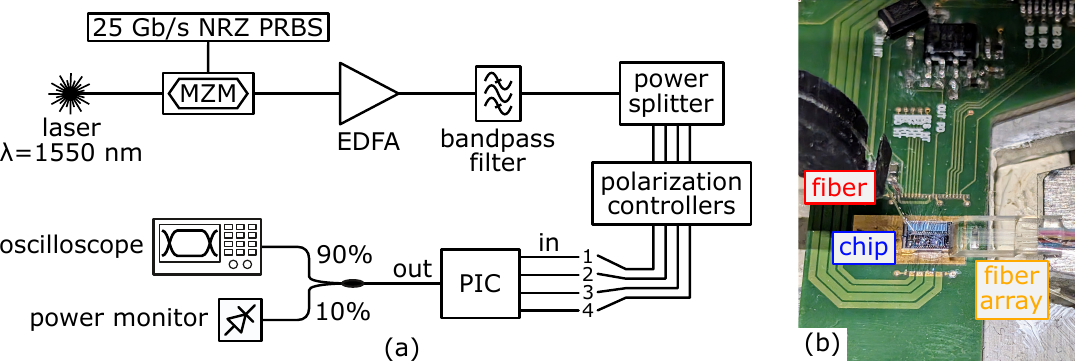}
	\caption{a) Schematic of the optical setup employed for validating the PIC functionality, the (de)multiplexers and the control strategy. b) Photograph of the PIC mounted and wire-bonded to the interface electronic board. All the fibers are single-mode.}
	\label{fig4}
\end{figure*}

The performance of the time-multiplexed control of the PIC implementing a multi-port coherent adder were tested with the optical setup shown in Fig. \ref{fig4}a. A laser source at \SI{1550}{\nano\meter} was modulated with a non-return-to-zero (NRZ) \SI{25}{\giga\bit\per\second} on-off keying (OOK) data stream to generate the input of the circuit. An erbium-doped fiber amplifier (EDFA) was used to compensate for the coupling losses at the chip input and output ports (\SI{4.5}{\decibel} per GC). The amplified spontaneous emission (ASE) noise of the EDFA was reduced with an ASE rejection filter, with a \SI{-3}{\decibel} bandwidth of \SI{0.25}{\nano\meter} centered around the carrier wavelength. The optical signal was equally split by a $1 \times 4$ fiber splitter and the polarization of the resulting beams were aligned to the transverse electric (TE) mode of the on-chip grating couplers before injecting them into the PIC with a fiber array. The output of the coherent adder was monitored with a single-mode fiber connected to an optical oscilloscope and a power monitor. 

The PIC was mounted on a custom interface printed circuit board (PCB) allowing easy optical coupling and stable connection to the electronic control board (Figure \ref{fig4}b). The board contains the front-end transimpedance amplifier needed to read the on-chip PDs and the power management circuits to supply the PIC. All other electronic functionalities needed to operate the control system are provided by a custom motherboard, connected through shielded cables to the interface PCB. The motherboard contains the ADC and DAC needed to convert the input/output control signals and the FPGA that executes the control algorithm and drives the MUX/DEMUX digital bits. The temperature of the PIC was measured with a \SI{10}{\kilo\ohm} thermistor placed on the interface board and, in the reported experiments, it is kept at \SI{28}{\celsius} with a thermo-electric cooler.

\subsection{Automated adaptive PIC configuration}\label{sec5_2}
As a first step towards the validation of the entire PIC, we started by testing the electro-optical response of every MZI. This test was carried out to verify the effectiveness of sequential PD readout (MUX) and heaters actuation (DEMUX and high-current drivers). Light was injected from input IN1 and the MZIs were tested sequentially till the last stage. Then, the sequence was repeated using the other inputs. For each MZI, the heater driver gate voltage was swept between 0 and \SI{10}{\volt} and the output was detected with the corresponding on-chip monitor PD, to retrieve the MZI transfer function. A \SI{50}{\milli\volt} sinusoidal dithering modulation at \SI{2}{\kilo\hertz} was superimposed to the heater voltage to measure also the derivative of the MZI transfer function, necessary for controlling the device \cite{zanetto_2021_dithering}. The full control procedure and signal processing pipeline are described in detail in the Methods section.

\begin{figure}[!t]
	\centering
\includegraphics[width=0.6\columnwidth]{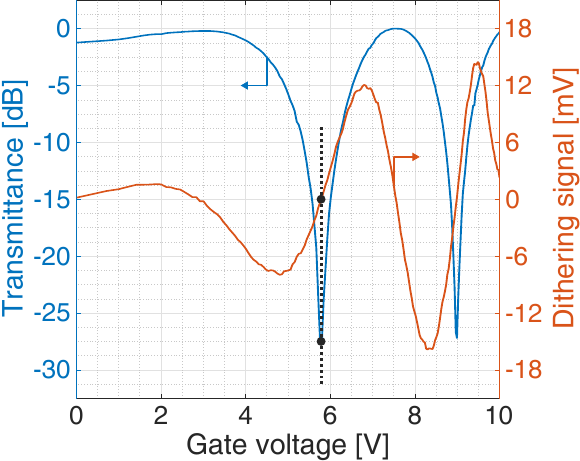}
	\caption{Measured transfer function of a MZI (blue) and its first derivative (red) obtained with the dithering technique. The stationary points of the MZI transfer function correspond to the zeros of the first derivative.}
	\label{fig5}
\end{figure}

The MZI transfer function and its derivative are reported in Fig. \ref{fig5} for one of the MZIs, all the others having very similar behaviour. Notice that, as expected, the measured dithering signal is zero in the stationary points of the transmittance characteristic. The slight increase of the transmission occurring from \SI{0}{\volt} to \SI{3}{\volt} is caused by a small native unbalance of the MZI due to fabrication tolerances. This unbalance randomly changes on the various MZIs, but it does not introduce any issues because it is automatically compensated by the implemented feedback control system. The test confirms that a $\pi$ shift is achieved at about \SI{5.8}{\volt} (corresponding to $\approx\SI{10}{\milli\watt}$ of heater dissipated power), where the extinction ratio (ER) is about \SI{27}{\decibel}. Notably, the current capability of the on-chip drivers can generate a phase shift of up to $3\pi$ at \SI{9}{\volt} ($\approx\SI{33}{\milli\watt}$). At the same time, Fig. \ref{fig5} certifies the precise temporal alignment of the multiplexing control signals, which do not impair the readout and actuation actions.

\begin{figure}[!t]
	\centering
    \includegraphics[width=0.6\columnwidth]{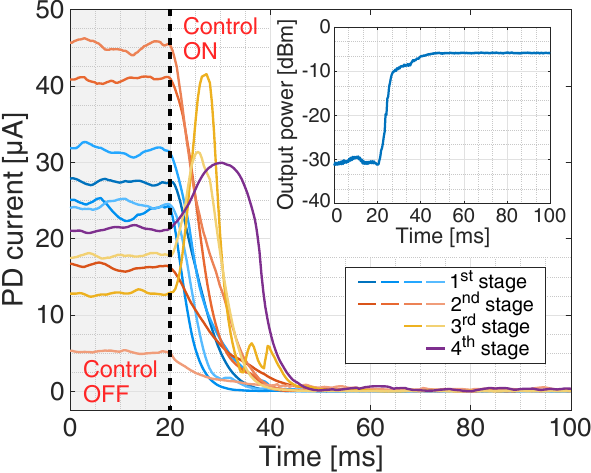}
	\caption{PDs current transient when activating the control loops from a random initial PIC configuration. The inset shows the corresponding evolution of the optical power at the chip output, which is maximized in about \SI{25}{\milli\second}.}
	\label{fig6}
\end{figure}

The calibration-free control algorithm was then tested by activating all the 14 electronic feedback loops at the same time to route and combine all 4 input beams into a single optical output. Figure \ref{fig6} shows the convergence transient of the PDs photocurrent during the automated configuration of the coherent adder. In order to demonstrate the robustness of the approach, the bias points of the MZIs were initially set by applying random voltages to the integrated heaters, and then the feedback loops were simultaneously activated. As expected, the control system correctly minimizes the PD photocurrents, thus routing and combining the four inputs to the output port of the coherent adder (see inset of Fig. \ref{fig6}). Each MZI is tuned in around \SI{10}{\milli\second}, allowing full configuration of the PIC in less than \SI{30}{\milli\second}.

\begin{figure}[!t]
	\centering
    \includegraphics[width=0.6\columnwidth]{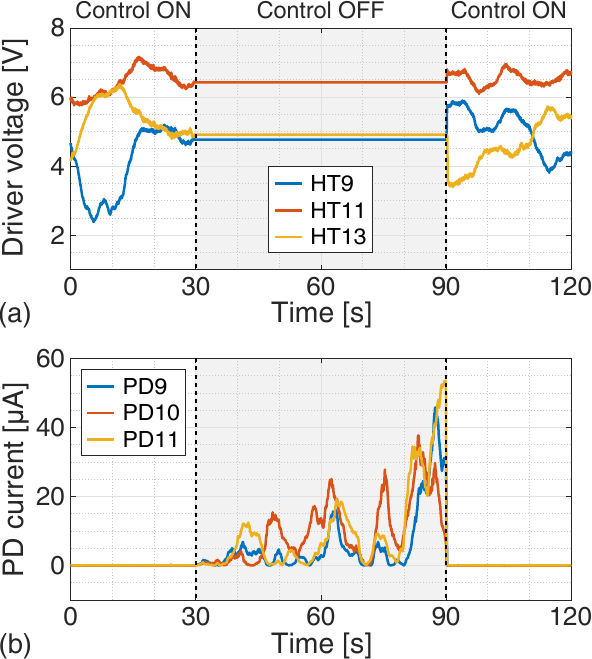}
	\caption{Temporal evolution of a) the heater driver voltages and b) PD currents when the control loops are enabled or disabled. When the feedback action is paused, the random phase fluctuations of the input signals impair their correct recombination by the coherent adder, as demonstrated by the oscillations of the PDs current. PD11 is the complement of the OUT signal.}
	\label{fig7}
\end{figure}

The dynamic feedback control is also able to counteract the effect of drifts and perturbations that might affect the photonic chip, like thermal fluctuations, acoustic noise and mechanical vibrations, as well as variations of the optical input signals. These phenomena, whose main contributions are concentrated below \SI{1}{\kilo\hertz} in short optical fiber connections, can cause random phase fluctuations of the input beams which, if not counteracted properly, would impair their coherent recombination by the PIC. We experimentally verified the stabilizing action of the control loops. Figure \ref{fig7} shows the temporal evolution of both the heater driver voltages (a) and the PD photocurrents (b) of the coherent adder, when the control is enabled or disabled. When activated, the feedback loops continuously update the actuators commands and keep all MZIs locked to the minima of their transfer function, as confirmed by the measured PD currents. Instead, when the control is disabled and the heater voltages are held at a constant value, the random input phase variations cause a partial recombination, with a time-dependent behavior dictated by the temporal dynamics of the external perturbations. In the considered example, such perturbations are mainly due to random phase drifts of the optical beams in the 4 fibers coupled to the photonic chip. The test shows that a static calibration of the mesh, based on a look-up table, is not sufficient to correctly recombine the input signals and highlights the importance of the active control and stabilization.

\subsection{\SI{25}{\giga\bit\per\second} communication demonstration}\label{sec5_3}
\begin{figure*}[!h]
	\centering
\includegraphics[width=\linewidth]{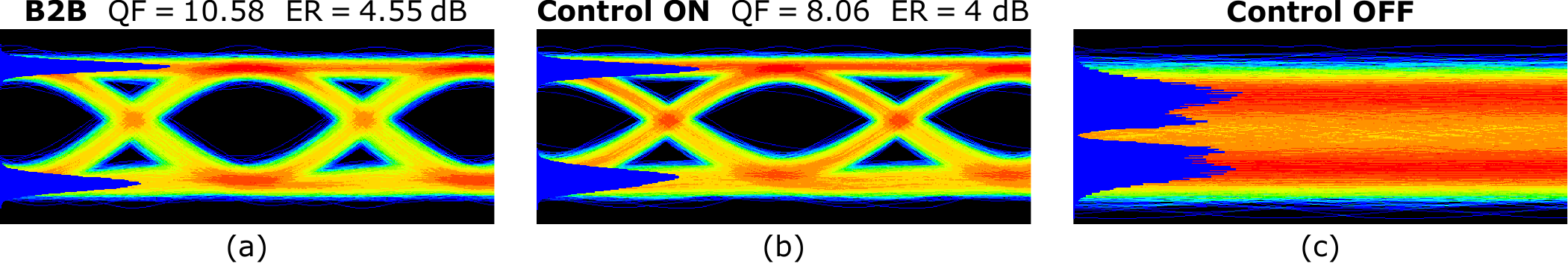}
	\caption{Eye diagrams of a \SI{25}{\giga\bit\per\second} signal measured at the output of the photonic chip, comparing its performance in several conditions. a) Back-to-back reference case, obtained by routing light through a straight waveguide with the same length as the PIC. b) Signal at the output of the coherent adder when the active control feedback is enabled or c) disabled. As expected, active stabilization is needed to recombine the input beams correctly.}
	\label{fig8}
\end{figure*}

The experimental validation of the dynamic phase locking achieved by the proposed time-multiplexed control was confirmed with a \SI{25}{\giga\bit\per\second} high-speed data transmission. Here, the quality of the output data signal is evaluated, simulating the operation in a real communication scenario. Figure \ref{fig8} illustrates comparative eye-diagram analyses of the signal at the output of the coherent adder under different operating conditions. The back-to-back reference measurement (B2B), reported in panel a) and obtained by transmitting the signal through a straight waveguide with the same geometrical length as the coherent adder, shows a quality factor (QF) of 10.58 and an ER of \SI{4.55}{\decibel}, representing the baseline performance. Panel b) reports the eye diagram when the 4 input beams are recombined by the coherent adder with the control enabled. In these conditions, the signal quality (QF $= 8.06$ and ER $= \SI{4}{\decibel}$) confirms good recombination of the four portions of the signal injected into the PIC, with only minimal degradation (approximately \SI{0.55}{\decibel} penalty) compared to the reference case. 
The measured Q-factor guarantees an error-free performance with a large margin.
Instead, when the control loop is deactivated, the random input phase fluctuations in the input fibers prevent correct combination of the four inputs, a completely closed eye diagram is observed and the input signal cannot be recovered. Such result confirms the fundamental role of dynamic phase locking when the PIC is operated as a coherent adder. 

\subsection{Phase sensing demonstration}\label{sec6}
Since the MZI heaters modify the phase of the propagating beams to achieve the desired coherent combination at the PIC output, their driving voltage contains information on the phase of the input signals. In other words, the continuous control action performed by the feedback electronics allows us to assess the input phase variations in real-time. This can be useful for estimating, among others, the direction of arrival of an optical beam or the magnitude of wavefront distortions caused by propagation through aberrators, scattering media, multimode or multicore fibers, or turbulent environments \cite{miller_2013_self,martinez_2024_selfadaptive,milanizadeh_2021_coherent,milanizadeh_2022_separating,miller_2013_self_2,annoni_2017_unscrambling}. This evaluation is possible if the phenomena of interest have a slower temporal evolution than the bandwidth of the control loops, so that the working point of each MZI can be constantly adjusted to perform a correct coherent combination.

\begin{figure}[!t]
	\centering
        \includegraphics[width=0.6\textwidth]{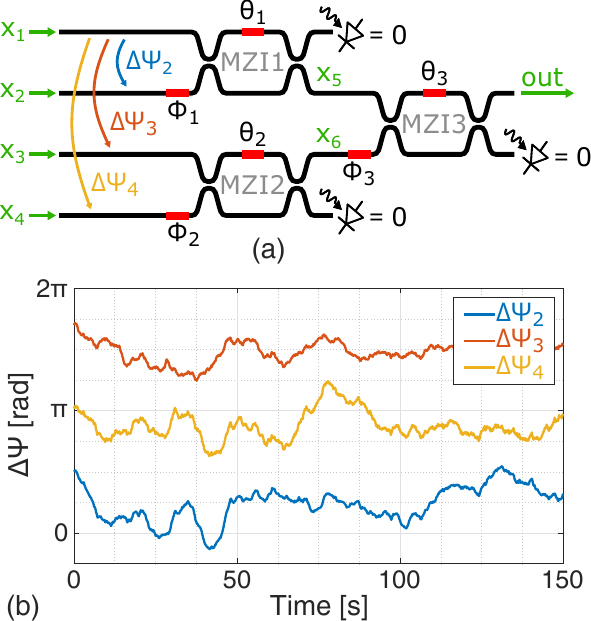}
	\caption{a) $4 \times 1$ binary tree of MZIs used to evaluate phase variations of the input beams. b) Experimental extraction of the input phase drifts, obtained by analyzing the command of the on-chip actuators and showing time-dependent variations in a time interval of 2.5 minutes.} 
	\label{fig9}
\end{figure}

To provide direct evidence, let us consider the 4-channel coherent adder shown in Fig. \ref{fig9}a, which represents the final stage of our PIC. The circuit inputs are four beams $x_i=\lvert x_i\rvert\cdot e^{\psi_i}$, where $\lvert x_i\rvert$ is the magnitude of the electric field and $\psi_i$ is its phase. To perform a correct coherent summation (that is, to null the signal read by the PD), the phase shifts introduced by the actuators of MZI1 must satisfy the following conditions:

\begin{equation}
    \begin{cases}
      \theta_1 = 2 \tan^{-1} \lvert \frac{x_1}{x_2} \rvert +\pi \\
      \phi_1 = \psi_2-\psi_1
    \end{cases}
    \label{eqn1}
\end{equation}
The phase shift $\phi_1$ induced by the first heater (proportional to its dissipated electrical power) provides the phase difference between the two inputs $x_2$ and $x_1$, while $\theta_1$ is related to the ratio of their amplitudes. The phase of the output signal $x_5$ is given by $\psi_5=\pi/2 - \theta_1/2 + \psi_1$. Similarly, coherent sum at the upper output port of MZI2 requires 

\begin{equation}
    \begin{cases}
      \theta_2 = 2 \tan^{-1} \lvert \frac{x_3}{x_4} \rvert \\
      \phi_2 = \psi_4-\psi_3
    \end{cases}
    \label{eqn2}
\end{equation}
and the phase of $x_6$ is $\psi_{6}= -\pi/2-\theta_{2}/2+\psi_3$. 

The same analysis can be repeated on MZI3. Heater $\phi_{3}$ cancels the phase difference between $x_5$ and $x_6$ and $\theta_{3}$ accounts for their amplitude ratio. For larger coherent adders, this operation can be repeated sequentially for all the MZIs until the output of the mesh is reached, without increasing the complexity of the calculation. The acquired information can then be used to evaluate the phase differences of each input beam with respect to $x_1$, as 

\begin{equation}
    \begin{cases}
      \Delta \Psi_2= \psi_2-\psi_1 = \phi_1 \\
      \Delta \Psi_3= \psi_3-\psi_1 = \pi -\theta_1/2 + \theta_2/2 + \phi_3 \\
      \Delta \Psi_4 = \psi_4-\psi_1 =  \pi -\theta_1/2 + \theta_2/2 + \phi_2 +\phi_3
    \end{cases}
    \label{eqn3}
\end{equation}
These relations can be easily extended in the case of a larger number of inputs. 

Figure \ref{fig9}b shows the experimental validation of this approach. The relative phase difference between the inputs of the adder has been extracted in real time by analyzing the voltages of heaters $h9$ to $h14$ of our PIC over 2.5 minutes. It is computed by knowing the thermal efficiency of the actuators, which allows us to convert their dissipated power into the corresponding induced phase shift. The measurement confirms the presence of phase fluctuations due to environmental effects, which can be as high as $0.5 \pi$ on a period of a few minutes for the case of the fiber array employed in our setup. A measurement accuracy of $\approx\pi/250$ has been extracted by computing the root-mean-square value of the noise superimposed on the heaters command. This result demonstrates that the dynamically-controlled coherent adder can be used for JCAS purposes and opens the way to new advanced applications for this photonic structure.

\section{Conclusions and discussion} \label{sec7}

We experimentally demonstrated the electronic control of reconfigurable silicon PICs with a fully on-chip time-multiplexed approach. This method supports both the readout of integrated light sensors and the actuation of volatile thermal phase shifters. The effectiveness of this technique was demonstrated by accurate and stable configuration of an integrated MZI-based architecture. Notably, the control strategy operates, at the same time, on devices that are not sensitive to the input signal phase (such as the 2-stage switch matrix) and on phase-sensitive components (such as the 4$\times$1 binary mesh topology implementing the coherent adder). With such control, the photonic chip simultaneously performs communication and sensing functionalities, successfully reconstructing a \SI{25}{\giga\bit\per\second} signal and estimating the phase difference between the input beams. 

The time-multiplexed approach offers a viable solution for scaling the feedback control paradigm to larger photonic chips, without correspondingly increasing the required electrical connections. These can represent a significant bottleneck, in terms of area occupation and reliability, when the number of on-chip devices increases above a few tens \cite{fowler_2022_integrated,ribeiro_2020_column}. However, multiplexing $N$ devices reduces the maximum control bandwidth that can be achieved by the same factor $N$, as compared to the fully parallel case. Therefore, when the chip scale increases significantly, a hybrid solution where several MUXes/DEMUXes are used in parallel, sharing the same digital selection bits, can be a viable option for maintaining sufficient control bandwidth while still reducing the number of electrical I/O lines.

Using the same PIC for both communication and sensing enables the extraction of relevant information about the transmitted beams and the optical link without complex post-processing algorithms, which typically require significant power consumption. Therefore, the design of circuits capable of providing this functionality is expected to become increasingly popular, further extending the range of applications of PICs. In the proposed case, the sensing capability of the coherent adder is inherently linked to the maximum control bandwidth that can be achieved, which determines the maximum phase tracking speed. This could be increased by choosing a higher dithering frequency, which is the bandwidth-limiting factor in our demonstration. While the MUX/DEMUX would support a higher switching rate to accommodate the higher dithering frequency \cite{zanetto_2023_unconventional}, the settling times of the acquisition and actuation chains of our external electronic platform imposed a practical limitation in the presented demonstration.  

The integration of electronic and photonic functionalities on the same die can be achieved using several technologies. Considering the requirements of our application, we chose to integrate electronic devices into a commercial silicon photonics platform with a zero-change approach \cite{zanetto_2023_unconventional}. Indeed, the required electronic circuits need to operate at a relatively slow speed to maintain closed-loop control of the optical functionality. Therefore, the use of advanced electronic-photonic technology nodes, although possible, would result in an increased cost without any real benefit coming from the scaled CMOS devices. Our approach only employs the conventional fabrication processing steps of commercial silicon photonic foundries, thus not requiring any post- or custom processing and being suitable for multi-project wafer runs, as in the case of the chip presented in this work. The use of a commercial foundry also demonstrates the compatibility of our design with industry-standard platforms.

\section*{Methods}

\subsection*{Digital electronics for feedback control} \label{sec:methods}
\begin{figure*}[!t]
	\centering
\includegraphics[width=\linewidth]{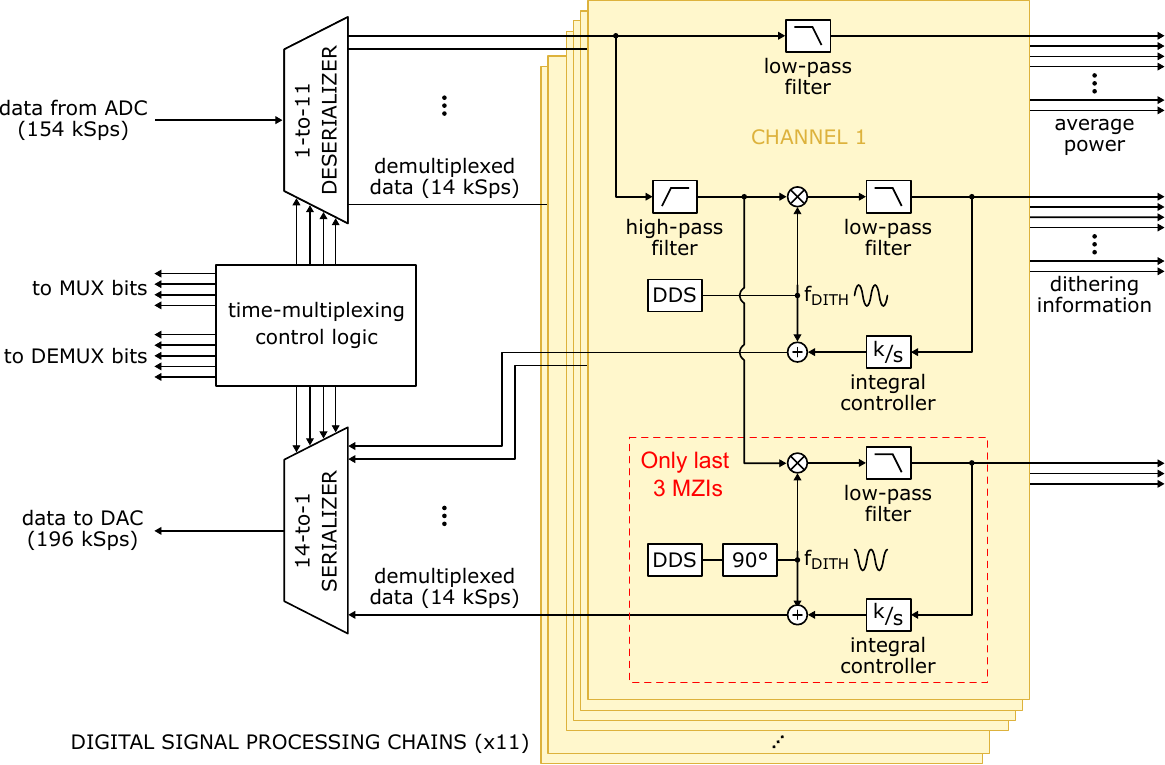}
	\caption{Schematic of the time-multiplexed digital control feedback loop employed for configuring and locking the optical functionality of the PIC.}
	\label{fig3}
\end{figure*}

The time-multiplexed approach is included in the digital feedback logic used to reconfigure the functionality of the optical circuit. Its schematic is reported in Figure \ref{fig3}. The average light power measured by each PD is obtained by simply low-pass filtering the readout after the DEMUX. A calibration-free strategy based on the dithering technique is instead used to configure and lock the working point of each MZI \cite{zanetto_2021_dithering}. The technique requires adding a small sinusoidal signal to the heaters voltage, thus slightly modulating the light intensity at the output of each MZI proportionally to the first derivative of their transfer function. This information is extracted with a digital lock-in readout and fed to an integral controller, which updates the heaters command until the measured dithering modulation is brought to zero. In this way, the MZIs are locked to their stationary points. We set the sign of the feedback loop to target the MZI minimum transmission points and route the input light towards the output of the adder.  

The same dithering frequency $f_{dith} = \SI{2}{\kilo\hertz}$ is used to control all the MZIs integrated in the photonic circuit. Indeed, once an MZI is tuned, the dithering oscillation at its output is brought to zero, therefore it does not impair the configuration of the following devices \cite{zanetto_2021_dithering}. The MZI switches integrate only one heater each, so they require a single digital processing chain, whereas the MZIs of the coherent adder need two parallel control loops to operate both heaters. In the latter case, to discriminate the effect of each heater from the readout of a single PD, we exploited the phase selectivity of the lock-in technique and used two in-phase/quadrature orthogonal dithering modulations at the same frequency, as shown in Figure \ref{fig3}. This simplifies the control procedure and the scaling of the proposed approach, since all the digital chains process the same kind of signals. 

The entire control logic is implemented in the FPGA (Xilinx Artix-7, mounted on a commercial Opal Kelly XEM7310 module) that is hosted by the external electronic board, without needing any co-processor. Indeed, the feedback algorithm has been chosen because it does not require any complex algebraic operation that cannot be handled by the FPGA, in addition to its calibration-free nature.

\section*{Statements and Declarations}

\subsection*{Acknowledgments}
This work was partially supported by the European Commission under the Horizon Europe Programme, Chips JU PL5 PIXEurope (Advanced Photonic Integrated Circuits Pilot Line for Europe, grant no. 101213727). The authors thank Polifab, the nanofabrication facility of Politecnico di Milano, for dicing and wire-bonding the chips.   

\subsection*{Author contributions statement}
S.D. and F.Z. developed the control electronics algorithm. F.Z. and M.C. designed the layout of the PIC. S.D. set up and carried out the experiments. S.D., F.M. and G.F. analyzed the results. A.M. and M.S. supervised the project. S.D., A.M., F.M, and F.Z. wrote the manuscript. All the authors contributed to the revision of the manuscript.

\subsection*{Data availability}
All the data supporting the findings of this study are available within this article. Any additional data are available from the corresponding author upon reasonable request. 

\subsection*{Competing interests}
The authors declare no competing interests. 

\bmhead{Funding} Chips Joint Undertaking, 101213727

\bibliography{bibliography}

\end{document}